\documentstyle[sprocl]{article}
\input{psfig}

\def\be{\begin{equation}}
\def\ee{\end{equation}}
\def\bea{\begin{eqnarray}}
\def\eea{\end{eqnarray}}

\begin{document}
\title{ANISOTROPIES IN THE COSMIC MICROWAVE BACKGROUND: THEORY}
\author{ SCOTT DODELSON }
\address{NASA/Fermilab Astrophysics Center,
P.O. Box 500, Batavia, IL 60510, USA}
\maketitle\abstracts{Anisotropies in the Cosmic Microwave
Background (CMB) contain a wealth of information about the past
history of the universe and the present values of cosmological parameters.
I ouline some of the theoretical advances of the last few years.
In particular, I emphasize that for a wide class of cosmological
models, theorists can accurately calculate the spectrum to better
than a percent. The specturm of anisotropies today is directly
related to the pattern of inhomogeneities present at the
time of recombination. This recognition leads to a powerful
argument that will enable us to distinguish inflationary models
from other models of structure formation. If the inflationary models
turn out to be correct, the free parameters in these models
will be determined to unprecedented accuracy by the upcoming
satellite missions.
}

\section{History}

The Texas Symposium on Relativistic Astrophysics was held in Chicago
ten years ago in 1986. David Wilkinson spoke about the
cosmic microwave background. He undoubtedly made the point that the 
CMB provides us with some of the best evidence for the Big Bang. 
There was no evidence (and there still is no evidence) for
any deviations from a black-body spectrum. And this is one of the 
primary predictions of the Big Bang. 

Wilkinson devoted most of
his talk to searches for anisotropies in the CMB. 
The fact that the CMB temperature is the same in all directions indicates
that the universe was very smooth early in its history. However, 
cosmologists generally work within the framework of gravitational
instability which says that small inhomogeneities early on 
grew via gravity into the large structures we see today. Thus, the CMB
should {\it not} be perfectly isotropic; it should carry
some imprint of those small, early inhomogeneities. Wilkinson 
compiled the upper limits on anisotropies from the experiments of the
time. This compilation is reproduced in Figure 1, where
 I have taken the liberty
of slightly changing his notation. In particular, it is convenient
to expand the temperature on the sky in terms of spherical harmonics
\begin{equation}
{T(\theta,\phi)\over T_0} = \sum_{l=0}^\infty \sum_{m=-l}^l a_{lm}
Y_{lm}(\theta,\phi).
\end{equation}
When we expand in this fashion, low $l$'s correspond to anisotropies
on large angular scales (the quadrupole is $l=2$) while large
$l$'s correspond to anisotropies on small scales. The square of the
coeffients of the $Y_{lm}$'s are known as the $C_l$'s. These are
extremely useful things becuase they can be calculated by
theorists and measured by observers. A given experiment at
angular scale $l$ measures $\delta T_{rms} \sim
[l(l+1)C_l/2\pi]^{1/2}$. The upper limits at the time correspond to
$\delta T_{rms} \sim 50-200 \mu$K. 

\begin{figure}[p] 
\centerline{\hbox{
\psfig{figure=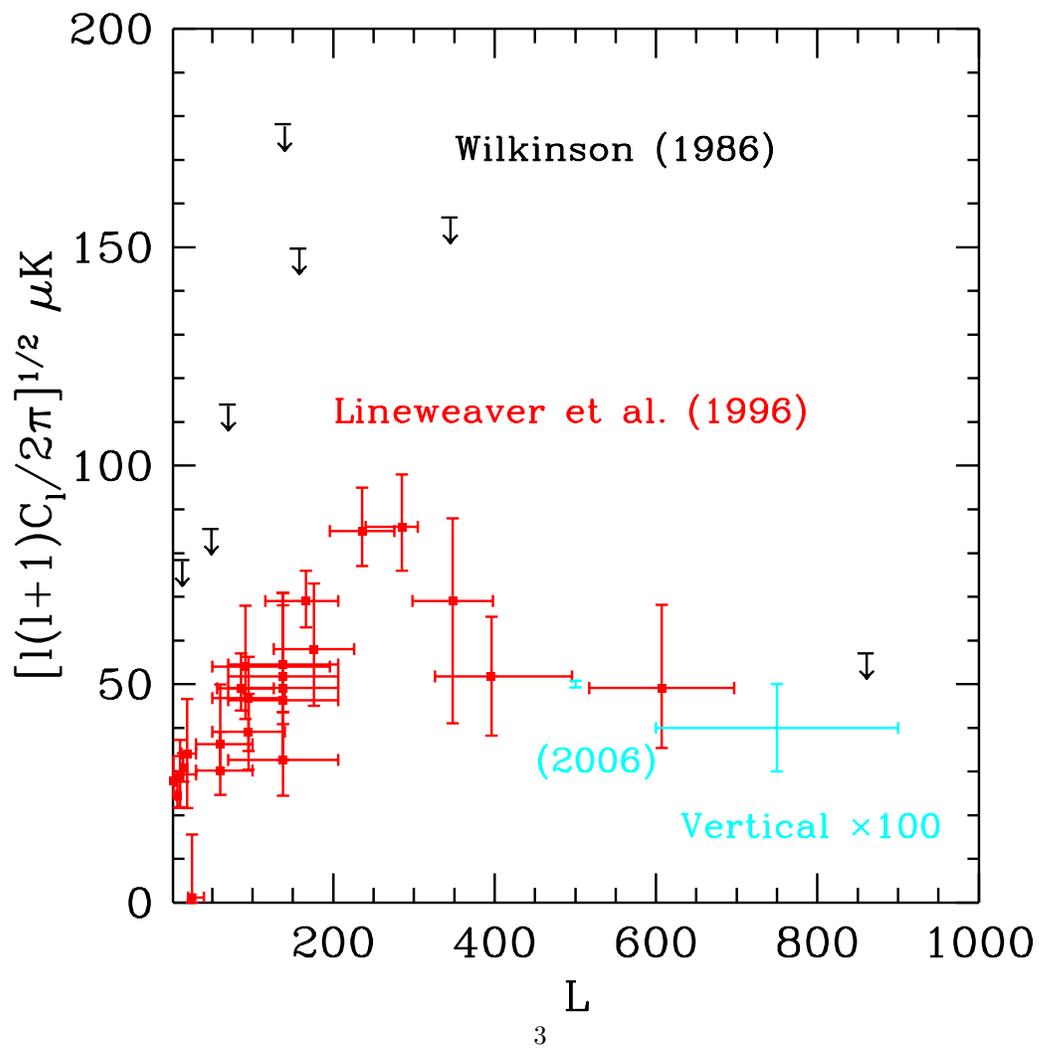,height=18cm,width=14cm}}
}
\caption{Observations of the CMB spectrum. Upper limits
are those compiled by Wilkinson in 1986. Detections
were compiled by Lineweaver based on results within the
last two years. Anticipated error bars from satellites
are also shown.}
\label{fig:cmb}
\end{figure}

Wilkinson was obviously aware of the fact that these upper limits
were tantalizingly close to the levels of anisotropies predicted
by many theories. He ended his talk by saying, ``{\it If the anisotropies
are indeed just below current limits, as most of us
feel they must be, the next few years should see this field turn from 
one of searching to one of studying.}''

\section{Experiments}

How has the field progressed since Wilkinson's review?
Figure 1 shows, along with Wilkinson's compilation,
a recent compilation\cite{LINE}\ 
of all experiments in the last
two years. Starting with the COBE\cite{COBE}\ 
detection in 1992
at the largest angular scales, there have been dozens
of detections on a wide range of angular scales. These
detections, as Wilkinson's quote makes clear, 
were anticipated based on typical models of structure
formation. Although the details are not yet in, it is
safe to say that gravitational instability theories
predicted the level of anisotropies that are observed today.

Another feature of the detections is just now becoming 
evident. As one moves from low $l$ to high $l$ (from
large scales to small scales), one sees evidence of 
a gradual rise in the the amplitude of the anisotropies.
We will see shortly that this too is a prediction of
some of the more popular models of structure formation.

How will the situation look ten years from now when
results from the current crop of balloon-borne
and ground-based experiments have come in, and the 
two satellite experiments (MAP\cite{MAP}\ and PLANCK\cite{PLANCK}) 
will have
made all-sky maps? Figure 1 shows the expected error
bars in the year 2006. There are several ways to represent
the knowledge we will have at the time. First, it is important
to note that, today, experiments are sensitive to a range of
$l$'s: thus, $C_{500}$ for example is not measured by a given
small scale experiment. Rather, each experiment measures
a signal integrated over a wide range of $l$. This range
is depicted by the horizontal error bars in Figure 1. 
In the future, we can continue to smooth over the $l$'s in
this fashion. Then the errors will be as shown on the far
right in Figure 1. In order to see them on this graph,
I have blown them up by a factor of 100! We will also
have the ability by then, though, to determine each
individual $C_l$. The expected errors on $C_{500}$ are shown
in Figure 1. Either way you look at it, we will have an extraordinary
amount of information in ten years. For the experimentalists,
at least, it is clear that Wilkinson's prediction has come true. The field
really has moved from searching to studying.

\goodbreak
\section{Theory}
\nobreak

CMB theorists have also been very active over the last
few years. First of all, for a wide range of models, we are
confident that we can calculate\cite{BE}\ the anisotropy spectra --
the $C_l$'s -- to an accuracy of better than a 
percent.
Figure 2 shows the results of seven different groups who
independently calculated the $C_l$'s for a given model. This
graph was made about two years ago, and the agreement has only
gotten better since then. Not only can we calculate accurately,
but we can also calculate quickly. Thanks to 
Seljak and Zaldariagga\cite{SELJAK}, in the time it has taken me to write this
paragraph, we could have run off another set of $C_l$'s.

\begin{figure}[t] 
\centerline{\hbox{
\psfig{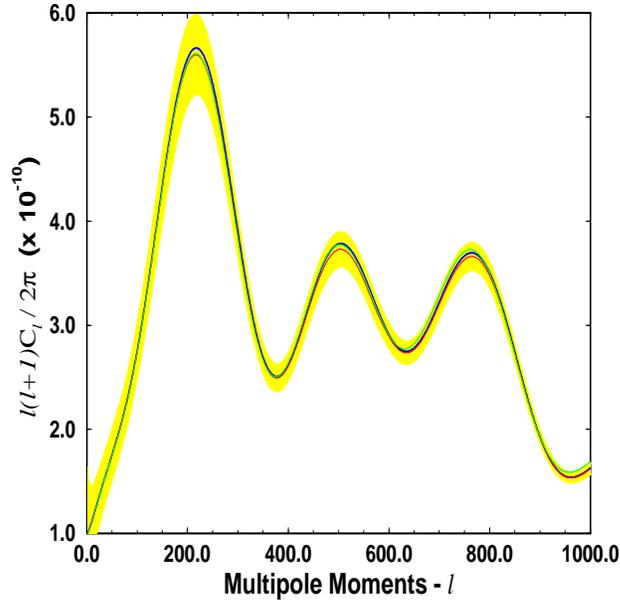}}
}
\caption{Seven different calculations of the CMB power spectrum
for a model with adiabatic fluctuations. All lie within the shaded
band which illustrates the minimum ``cosmic variance'' errors.}
\label{fig:NOW}
\end{figure}

We also have made great strides in the last few years
understanding the bumps and wiggles in the theoretical
curves. To understand the structure of these anisotropies,
we need to review the thermal history of the universe.
Recall that, early in the history of the universe, the temperature
of the cosmic gas was very high. So, anytime a free electron
and proton came together to form a hydrogen atom, a high energy photon
immediately destroyed it. There was essentially no neutral
hydrogen early on. This situation changed dramatically when the
temperature dropped below $1/3$ eV. After that time, there
were not enough ionizing photons around. So almost all
the free electrons and protons combined into neutral hydrogen.
This had dramatic implications for the cosmic photons. As long
as the electrons were free, they interacted with the photons
via Compton scattering. After they combined into hydrogen,
the photons travelled freely from the ``surface of last scattering''
to us today. So, for the purposes of the CMB, the universe
is neatly divided into two epochs: Before Recombination when
the photons and electrons behaved as a tightly coupled
fluid and After Recombination when photons freestreamed. 
The mathematics of freestreaming is a little complicated, but
the physics is completely trivial: it just requires us to 
trace the paths of free photons. So the physics behind the spectrum
of anisotropies comes solely from the epoch
Before Recombination. 

It pays to reiterate that Before Recombination, the photons and electrons
acted as a {\it fluid}. By this, I mean that it could be described
by only its $l=0$ component (as opposed to all the multipole
moments that are needed to describe it today). This represents an immense
simplification: instead of solving a infinite heirarchy of coupled
differential equations for all the photon moments, we need solve
for only one of the moments. The forces acting on this moment,
let's call it $\delta T$, are pressure and gravity. These forces
act in opposite directions. Pressure tends to smooth out
any inhomogeneities (i.e. drives $\delta T$ to zero) while gravity
produces inhomogeneities. It is not surprising then that acoustic
oscillations are set up in the medium. In fact, 
Hu \& Sugiyama\cite{HUSUG}\ have
shown that this oscillation pattern is precisely the one imprinted
in the $C_l$ spectrum of Figure 2. A quantitative analysis shows that
there are two possible modes\footnote{The modes look this
simple only in the idealized case of zero baryons and pure
matter domination. Accounting for baryons and other 
complications though does not alter
the qualitative fact that there are two very distinct
modes.} that can be excited in this fluid.
In particular, 
\begin{equation}\label{MODES}
\delta T(\vec x,\eta) = \int d^3k  e^{i\vec k\cdot\vec x}
\left[ A \cos[k\eta/{\sqrt3}]
+ B \sin[k\eta/{\sqrt3}] \right]
\end{equation}
where $\eta$ is conformal time.
Again, not surprisingly, the $C_l$ spectrum today is radically
different if the sine mode is excited than if the cosine mode 
is excited. Figure 3 shows that, as you would expect, the spectra are out
of phase with each other. 

\begin{figure}[t] 
\centerline{\hbox{
\psfig{figure=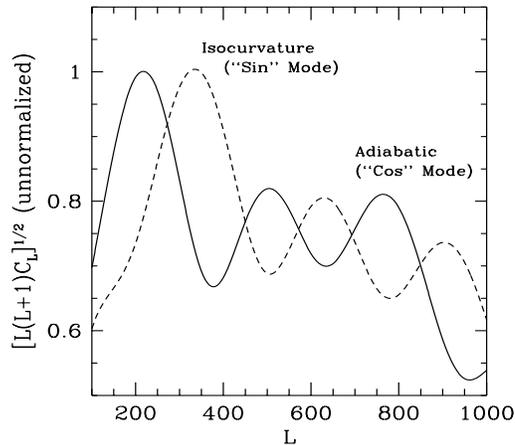,height=8cm,width=7cm}}
}
\caption{The power for two different theories, one of which (adiabatic)
excites the cosine mode of acoustic oscillations the other (isocurvature)
the sine mode. Inflationary predictions typically look like the adiabatic
spectrum here. Defect models should have some of the features of
the isocurvature spectrum, but the calculations at present are not yet
believable.}
\label{fig:myiso}
\end{figure}

It is clear from the present data shown in 
Figure 1 that we will shortly
be able to tell which of the two theoretical curves in
Figure 3 is more accurate. That is,
we will soon know whether the sine or the cosine mode were
excited in the early universe. This is extremely important
because we expect the two most popular mechanisms of structure
formation -- inflation and topological defects -- to excite different
modes. Let me walk through this argument which has
recently been clearly elucidated by Hu \& White\cite{HUWHITE}. 
Any theory which respects
causality necessarily requires that there be no correlations
on very large scales (scales that have not been in causal
contact with each other). This is equivalent to a boundary condition
on $\delta T$; namely that the Fourier transform vanishes at $k=0$.
This means that only $B$ in equation \ref{MODES}\ can be non-zero.
So topological defects, which of course obey causality, can be expected
to excite the sine mode. Inflation is a theory which introduces correlations
amongst scales that appear to be causally disconnected. 
Thus, inflationary models
can, and most often do, excite the cosine mode. 

\begin{figure}[t] 
\centerline{\hbox{
\psfig{figure=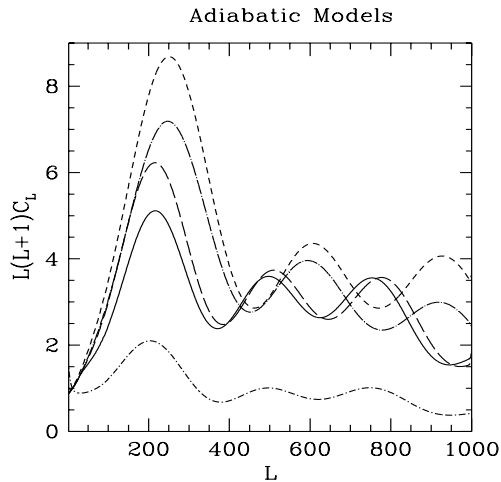,height=8cm,width=7cm}}
}
\caption{The spectra for adiabatic models with different
sets of parameters. Varied are the cosmological constant,
neutrino mass, spectral index, and Hubble constant.}
\label{fig:adiab}
\end{figure}

There are several caveats to the above argument. First of all,
the predictions for the adiabatic models depend on 
various cosmological parameters\cite{COSCON}: 
the slope of the primordial spectrum,
the contribution from tensor modes, the Hubble constant, the
baryon density, and several others. Thus the actual curves share
some of the  features of the curve labelled
``Adiabatic'' in figure 3, but the predictions are
by no means unique (see figure 4). Fortunately there are some robust
features of these curves which hold up even after allowing
many parameters to vary. The second caveat is that we simply
do not know for sure that defect theories follow the general isocurvature
model. There have been a few calculations of the 
spectrum in defect models\cite{TUROK}.
As one who is actively at work on one such calculation,
I think it is fair to say that we have not yet reached agreement.

Assuming there are no major theoretical surprises, we
can expect the experiments over the next several
years to pick out whether toppological defects or inflation
are correct. Once that issue is settled, it remains
to pin down the cosmological parameters which impact upon
the spectrum. One might think that since there are so many
free parameters, they cannot all be determined simultaneously.
Recent work has shown that this is {\it not} true. Figure 5 shows
an example\cite{WILL}: we let five parameters vary and show the
error ellipses projected down onto a couple of
two dimensional planes. The top figure shows that it
is quite possible that by the year 2006, we will not be arguing
about whether the Hubble constant is $50$ or $100$, but rather
whether it is $50.5$ or $50.0$. The bottom figure shows that,
in addition to the cosmological parameters, we should get a good
handle on the inflationary parameters, thereby allowing us
to distinguish amongst different inflationary models\cite{KT}.
A number of groups\cite{JUNG}\ have varied even
more parameters and all have reached the same general 
conclusion: the cosmological parameters will be pinned down
to unprecedented accuracy by the satellite experiments.

\begin{figure}[p]
\centerline{\hbox{
\psfig{figure=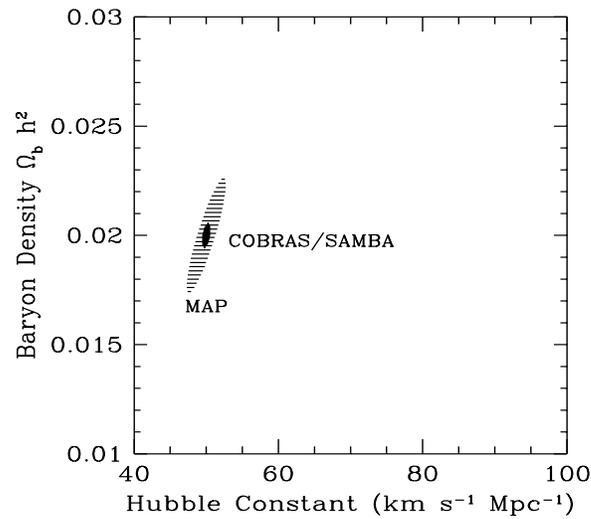,height=9cm,width=8cm}}
}
\centerline{\hbox{
\psfig{figure=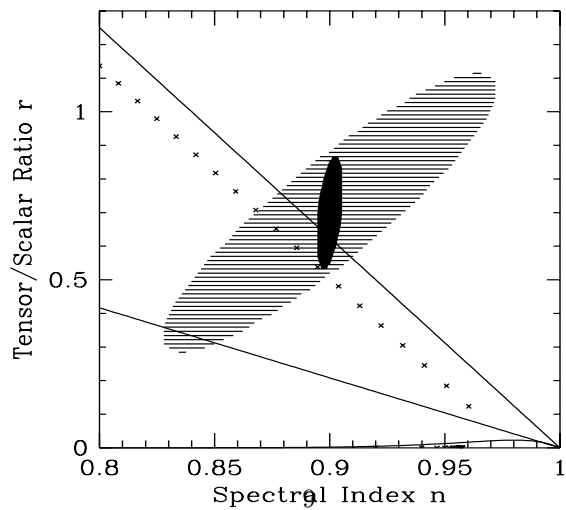,height=9cm,width=8cm}}
}
\caption{The estimated 95\% contours for the MAP (larger
ellipses in each case) and PLANCK
(formerly COBRAS/SAMBA) satellites. In each case five variables
-- the amplitude of the scalar and tensor perturbations, 
the spectral index of the scalars, the baryon density, and the Hubble
constant -- are allowed to vary. The ellipses are the projections of
the five dimensional ellipses onto the (Hubble constant,Baryon density)
plane and the (spectral index,tensor/scalar ratio) plane. 
The points and lines in
the $n-r$ plane correspond to the predictions of different inflationary
models.}
\label{fig:nr}
\end{figure}

\section{Conclusions} About thirty years ago, Penzias and Wilson
discovered the cosmic microwave radiation. This discovery convinced the
vast majority of physicists that the Big Bang model was correct. 
In 1992, the COBE satellite discovered anisotropies in the CMB. The
existence and amplitude of these
anisotropies were {\it predicted} by theories 
which relied on gravitational instability to form structure. 
It is perhaps too
early to know for sure, but I would guess that COBE's most enduring legacy
will be its evidence that current models of structure 
formation are on the right track. 

A number of cosmologists are beginning
to speculate about what we will have learned in ten years 
after the next generation of balloon and ground based experiments
and after the MAP and PLANCK satellites have flown. There is a very good
chance these measurements will clearly distinguish between the two most
popular models of structure formation: inflation
and topological defects. Indeed, this could happen very soon. If a peak
does indeed develop in the $C_l$ spectrum at around $l\sim200$, this will
be strong evidence for inflation. If the
general picture of inflation is
verified in this manner, the fun will begin. It will then be possible
to determine many of the cosmological parameters to unprecedented accuracy.
Further, the experiments will contain so much information that it will
be possible to distinguish amongst different inflationary models. 
This opens a window to study physics at energies that are
twelve orders of magnitude higher than those probed by 
the largest accelerators.

Many people are fond of pointing out that 
something completely unexpected
and confusing may turn up, thereby upsetting the possibility
of any such determinations. 
Of course this is possible. Unexpected and confusing discoveries
have rocked cosmology for decades. 
The ``man-bites-dog'' story in cosmology though is 
the one
in which the confusion ends; this may well happen within the
next ten years.

\section*{Acknowledgments} This work was supported in part by 
DOE and NASA grant NAG5--2788 at Fermilab.
.


\section*{References}
\begin{thebibliography}{99} 
\bibitem{LINE} C. H. Lineweaver {\it et al}, {\em astro-ph/9610133}.
\bibitem{COBE}G.P. Smoot {\it et al}, {\em Astrophys. J.} {\bf 396},
L1 (1992).
\bibitem{MAP} The MAP home page is
{\tt http://map.gsfc.nasa.gov/}.
\bibitem{PLANCK} The PLANCK (formerly COBRAS/SAMBA) home page is
{\tt 
http://astro.estec.esa.nl/SA-general/Projects/Cobras/cobras.html}.
\bibitem{BE} P.J.E.~Peebles and J.T.~Yu,  
{\em Astrophys. J.} {\bf 162}, 815 (1970); 
M.L.~Wilson and J.~Silk, {\em Astrophys. J.} {\bf 243}, 14 (1981);
J.R.~Bond and G.~Efstathiou, {\em Astrophys. J.} {\bf 285}, L45 (1984).
\bibitem{SELJAK} U. Seljak and M. Zaldariagga, {\em Astrophys. J.}
{\bf 469}, 437 (1996).
\bibitem{HUSUG} W. Hu and N. Sugiyama, {\em Astrophys. J.} {\bf 444},
489 (1995); F. Atrio-Barandela and A. G. Doroshkevich,
{\em Astrophys. J.} {\bf 420}, 26 (1994); P. Nasel'skij and
I. Novikov, {\em Astrophys. J.} {\bf 413}, 14 (1993);
U. Seljak, {\em Astrophys. J.} {\bf 435}, L87 (1994);
A. G. Doroshkevich. Ya. B. Zeldovich, and R. A. Sunyaev,
{\em Sov. Astron.} {\bf 22}, 523 (1978);
A. G. Doroshkevich, {\em Sov. Astron. Lett.}
{\bf 14}, 125 (1988).
\bibitem{HUWHITE} W. Hu and M. White, {\em Phys. Rev. Lett.}
{\bf 77}, 1687 (1996); N. G. Turok, astro-ph/9607109 (1996).
\bibitem{COSCON} J. R. Bond, R. Crittenden, R. L. Davis, G.
       Efstathiou and P. J. Steinhardt, {\em Phys. Rev. 
Lett.} {\bf 72}, 13 (1994).
\bibitem{TUROK} R. G. Crittenden and N. G. Turok, {\em
Phys. Rev. Lett.}
{\bf 75}, 2642 (1995); A. Albrecht, D. Coulson, P. Ferreira, and
J. Magueijo,  {\em Phys. Rev. Lett.} {\bf 76}, 1413 (1996).
\bibitem{WILL} S. Dodelson, W. Kinney, and E. W. Kolb, (1997).
\bibitem{KT} L. Knox and M. S. Turner, {\em Phys. Rev. Lett.}
{\bf 73}, 3347 (1994).
\bibitem{JUNG} L. Knox, {\em Phys. Rev.} {\bf D52}, 4307 (1995);
G. Jungman, M. Kamionkowski, A.
Kosowsky , and D. N. Spergel, {\em Phys. Rev.} {\bf D54}, 1332 (1996);
S. Dodelson, E. I. Gates, and A. S. Stebbins, 
{\em Astrophys. J.} {\bf 467}, 10 (1996).

\end{thebibliography}
\end{document}